\documentstyle[epsfig,12pt]{article}
\newcommand{\smallfrac}[2] {\mbox{$\frac{#1}{#2}$}}
\newcommand {\eqref} [1] {(\ref {#1})}
\newcommand {\beq} {\begin{equation}} 
\newcommand {\eeq} {\end{equation}}
 \newcommand {\ber}{\begin{eqnarray*}}
 \newcommand {\eer} {\end{eqnarray*}}
\newcommand {\bea}{\begin{eqnarray}}
 \newcommand {\eea} {\end{eqnarray}}

\def\Acknowledgements{\bigskip  \bigskip {\begin{center} \begin{large}
             \bf ACKNOWLEDGEMENTS \end{large}\end{center}}}

\begin{document}\begin{titlepage}
\rightline{CERN-TH/2001-272}
\vskip 1cm
\centerline{{\Large \bf UV/IR Mixing via Closed Strings}}
\centerline{{\Large \bf and Tachyonic Instabilities}}
\vskip 1cm
\centerline{Adi Armoni\ ${}^\dagger$ and Esperanza Lopez\ ${}^\ddagger$}
\vskip 0.5cm
\centerline{${}^\dagger$ Theory Division, CERN}
\centerline{CH-1211 Geneva 23, Switzerland}
\vskip 0.3cm
\centerline{adi.armoni@cern.ch}
\vskip 0.5cm
\centerline{${}^\ddagger$ Max-Planck-Institut f{\"u}r Gravitationsphysik, Albert-Einstein-Institut,}
\centerline{Am M{\"u}hlenberg 1,D-14476 Golm, Germany}
\vskip 0.3cm
\centerline{lopez@aei-potsdam.mpg.de}
\vskip 1cm

\begin{abstract}
We discuss UV/IR mixing effects in non-supersymmetric
non- commutative $U(N)$ gauge theories. We show that the singular (non-planar)
terms in the 2- and 3-point functions, namely the poles and the logarithms,
can be obtained from a manifestly gauge invariant effective action.
The action, which involves open Wilson line operators, can be derived
from closed strings exchange between two stacks of D-branes. Our
concrete example is type 0B string theory and the field theory that 
lives on a collection of $N$ electric D3-branes. We show that one of the
closed string modes that couple to the field theory operator
which is responsible for the infrared poles, is the type 0 tachyon.

\end{abstract}
\end{titlepage}

\section{Introduction}

Non-commutative gauge theories attracted recently a lot of attention,
mainly due to the discovery of their relation to string/M theory
\cite{Connes:1998cr,Seiberg:1999vs}. The perturbative
dynamics of these theories is very interesting:
planar graphs of non-commutative theories are exactly the same as
the planar graphs of ordinary theories apart from global phases which
depend on external momenta \cite{Gonzalez-Arroyo:1982ub,Gonzalez-Arroyo:1982hz,Filk:1996dm}. Non-planar graphs, on 
the other hand,  are regulated by the non-commutativity parameter 
$\theta$ and they are therefore UV-finite. This regularization is however 
only effective when there is a non-zero momentum inflow into the graph.
In particular, as a result of this, the non-planar contribution to the
propagator contains, usually, a pole $1/ (\theta p)^2$. This pole,
which originates from the high momentum region of the integral (UV) seems
to affect the large distance dynamics. This unusual phenomenon is
called UV/IR mixing \cite{Minwalla:2000px}. In supersymmetric theories this
pole cancels and
a softer version of UV/IR mixing exists due to a logarithmic
contribution \cite{Matusis:2000jf}. Aspects of the UV/IR mixing
phenomena in scalar theories
\cite{Arcioni:2000bz,Arefeva:2000uu,Micu:2001hp,Chepelev:2001hm,Griguolo:2001ez,Kinar:2001yk}
as well as gauge theories \cite{Hayakawa:1999zf,Bilal:2000bk,Armoni:2001xr,
Landsteiner:2000bw,Martin:2001bk,Pernici:2001va,Khoze:2001sy,Ruiz:2001hu,
Zanon:2001nq,Landsteiner:2001ky,Armoni:2001md} were studied by many authors
over the past two years. See \cite{Douglas:2001ba,Szabo:2001kg} for comprehensive reviews.

In this work we would like to focus on non-supersymmetric
non-commutative gauge theories in 4-dimensions. The non-planar pole 
modifies the dispersion relation of the photon as follows 
\cite{Matusis:2000jf,Ruiz:2001hu,Landsteiner:2001ky}
\beq
E^2= {\vec p\,}^2 - (N^{adj}_B-N^{adj}_F){g^2\over \pi^2}{1\over
(\theta p)^2}, \label{disp}
\eeq
where $N^{adj}_B$ and $N^{adj}_F$ are the numbers of bosons and
fermions in the adjoint representation, respectively. 
In the case of pure Yang-Mills theory, or in general when 
$N^{adj}_B>N^{adj}_F$, the low momentum end of the spectrum acquires
imaginary energy. Namely, the one loop analysis suggests that the
theory suffers from an instability. In \cite{Ruiz:2001hu} it was shown 
that the quadratic pole-like infrared divergence is
gauge-fixing-independent. In \cite{Landsteiner:2001ky} non-commutative
${\cal N}=4$ Yang-Mills at finite temperature was considered. This 
theory presents a regularized version of UV/IR mixing, where the 
temperature acts as a UV cut-off. Although perturbation theory
seems to be under better control in this case, tachyonic excitations at 
long wavelength appear for $T$ bigger than a critical temperature 
$T_c \sim 1/\sqrt{ g \, \theta}$. All this suggests that the associated 
instability might not be an artefact of perturbation theory, but 
instead its source should be searched in an expansion around 
the wrong vacuum. Of course, the existence of a stable vacuum is 
an open question.

As already mentioned, apart from the tachyonic poles in the non-planar
graphs, these graphs contain also subleading logarithmic contributions. 
These contributions do not change the photon dispersion relation and are 
not expected to cause any instabilities, but they are relevant in the IR. 
In particular, they affect the effective coupling 
\cite{Khoze:2001sy}. 
 
The purpose of this note is to gain a better understanding of the
tachyonic poles and the logarithmic effects by using string theory.
We summarize briefly our main results.
We consider a field theory that lives on 
D3 electric branes of the non-supersymmetric type 0B string theory. We
show that all singular amplitudes involving pole-like infrared
divergences can be encoded in a rather simple gauge-invariant effective 
action
\beq
S^I _{eff} \sim (N_B ^{adj} - N_F^{adj}) \int {d^4 p \over (2\pi)^4} 
\, {\rm tr}\, W(p)\ {\rm tr}\,  W(-p) \; {m^2\over \tilde p^2} 
K_2(m\tilde p) \, , \label{mainresult1}
\eeq
where $W(p)$ denotes the open Wilson loop operator 
\cite{Ishibashi:2000hs,Gross:2000ba}. 
This effective action is structurally analogous to
a closed string exchange between two stacks of D-branes. This coincidence 
is more than formal since we will see that the Bessel function kernel in 
\eqref{mainresult1} can be directly related to a 
closed string propagator in type 0 string theory.
Although \eqref{mainresult1} has the form of a closed string exchange,
all the tower of closed string modes contribute to it,
similarly to the proposal by \cite{Arcioni:2000bz}. Among the
closed string modes that couple to the open Wilson line operator
is the type 0 closed string tachyon. This is in contrast to the 
ordinary commutative case, where the tachyon just contributes to the 
vacuum energy of the field theory. The fact that the closed string 
tachyon couples to a non-trivial operator in the field theory, which
in addition is responsible of the pole-like infrared divergences, 
suggests that there could be a relation between string and non-commutative 
instabilities.

Although at a more intuitive level, our analysis can be extended
to the logarithmic infrared-divergent terms. They admit the following
gauge-invariant completion
\bea
& &
S^{II} _{eff} \sim \beta _0 \int {d^4 p \over (2\pi)^4} \left 
\{ {\rm tr}\, F_{\mu \nu}
W\,(p) \ {\rm tr}\, F_{\mu \nu} W \, (-p) \ + \right . \nonumber \\
& & 
\;\;\;\;\;\;\;\;\;\;\;\;\;\;\;\;\;\;\;\;\;\;\;\;\;\;\;\;\;
 \left . {\rm tr}\, D_\mu
\phi _i W\,(p) \ {\rm tr}\, D_\mu \phi _i  W\,(-p) \right \} K_0(m\tilde
p) \, . \label{mainresult2}
\eea
This expression is only schematic; a precise
definition of $S^{II}_{eff}$ is presented in section 4.
From a string point of view, we can interpret \eqref{mainresult2}
as due to the exchange of massive 2-form closed strings.
Note that this sector of the closed string does not contain tachyons. 
Indeed, as already mentioned, there is no tachyonic instability 
associated with the logarithmic part of the action. 
 
The organization of this article is as follows: in section 2 we
describe our model and we calculate the various singular amplitudes.
Sections 3 is devoted to a derivation of the full gauge invariant
effective action related to the infrared pole-like terms
in the case of pure $U(1)$ theory. 
In section 4 we suggest a derivation of
the effective action via closed strings exchange between D3-branes of
type 0 string theory. We discuss our results in section 5.

We use the following notations and conventions.
The field theory under consideration is a 4d one with space-space 
non-commutativity $[x^1,x^2]=i\theta$ (or a tensor 
$\theta ^{\mu \nu}$ with non-vanishing components in the
$1,2$ directions). We also use the notation $\tilde p^\mu = 
\theta ^{\mu \nu} p_\nu$. The $U(N)$ generators are normalized such that 
${\rm tr}\ t^A t^B = \smallfrac{1}{2} \delta ^{AB}$ and in particular 
the $U(1)$ generator is $t^0 = {1\over \sqrt{2N}}$.

Note added: as we finished our work, paper \cite{MVR} appeared.
The author of this paper arrived to the result \eqref{mainresult1} and
discussed it from the matrix theory perspective.

\section{Field Theory Calculations - \newline Various Non-Planar Amplitudes}

In this section we describe the UV/IR mixing effects in a concrete model. The
4d field theory under consideration is the theory that lives on
a stack of $N$ coincident electric D-branes of type 0B string theory.
It is obtained by dimensional reduction of pure (non-supersymmetric)
10d non-commutative Yang-Mills theory. The model contains a vector
and 6 adjoint scalars and it is described by the following action

\beq
S = {\rm tr}\ \int d^4 x \left( -{1\over 2g^2} F _{\mu \nu} \star 
F^{\mu \nu} + D_ \mu \phi ^i \star D ^\mu \phi ^i \right)
\eeq
where,
\beq
F_{\mu \nu} = \partial _\mu A_\nu - \partial _ \nu A_\mu - i( A_\mu
\star
A_\nu - A_\nu \star A_\mu)
\eeq
and
\beq
D_\mu \phi ^i = \partial _\mu \phi ^i - i (A_\mu \star \phi ^i - \phi^i \star
A_\mu)
\eeq
for $i=1...6$. The model is invariant under the following non-commutative
$U(N)$ gauge transformation
\bea
& & 
\delta _\lambda A_\mu = \partial _\mu \lambda -i (A_\mu \star \lambda -
\lambda \star A_\mu) \\
& &
 \delta _\lambda \phi ^i = -i (\phi ^i \star \lambda - \lambda \star
\phi ^i) .
\eea

Let us focus on the one-loop structure of the theory. The planar
sector is well understood. Apart from global phases associated to
external legs, the various amplitudes are the same as in the 
commutative cousin of the theory \cite{Filk:1996dm}. In particular the theory is one-loop
renormalizable with the same counterterms as those of the commutative
theory \cite{Armoni:2001xr}.

The non-planar sector of the theory exhibits an interesting
 pattern.
In this case the Moyal phases associated with the vertices do not 
cancel and lead to a UV-finite result \cite{Minwalla:2000px}. Consider
 first the
propagator of the gluon. The only non-vanishing non-planar graph
exists when the external gluons are in the $U(1)$
 \cite{Minwalla:2000px,Armoni:2001xr}. The sum of the
various contribution, due to gluons, ghosts and scalars running in the
loops yields \cite{Hayakawa:1999zf,Matusis:2000jf,Bilal:2000bk,Armoni:2001xr}
\beq
A ^{\mu \nu} _{(1-1)}= -8g^2 N \int {d^4 q \over (2
\pi)^4} {(2q^\mu q^\nu - g^{\mu \nu}q^2) \over q^4} \exp {2i\tilde p q}
= {64g^2 N \over (4\pi)^2} {\tilde p^\mu \tilde p^\nu \over \tilde p^4} .     
\label{vector}
\eeq
A similar calculation yields a similar result for the scalar
propagator
\beq
A _{(1-1)} = 8g^2 N \int {d^4 q \over (2
\pi)^4} {1\over q^2} \exp {2i\tilde p q}
= {32g^2 N \over (4\pi)^2} {1\over \tilde p^2} .
\label{scalar}
\eeq
Note that there is a relative factor of 2 between \eqref{vector} and
\eqref{scalar}. 

The poles in \eqref{vector} and \eqref{scalar} signal the interesting 
UV/IR mixing that is typical of non-commutative theories. The origin
of these contributions is the UV regime and they seem to affect
 the IR of the theory. These poles create a potential problem in the
renormalization process, since when the non-planar graphs are
inserted in higher loop diagrams they create new divergences.
 It was suggested \cite{Minwalla:2000px} that, in
certain cases, the sum of the geometric series of these contributions
can shift the pole such that these new divergences are avoided. This procedure,
however, cannot be implemented in the present case, due to the positive
sign in front of \eqref{vector} and \eqref{scalar}
\cite{Ruiz:2001hu,Landsteiner:2001ky}. In general, the coefficient in
front of \eqref{vector} and \eqref{scalar} is
determined by the number of bosons in the adjoint representation minus
the number of fermions in the adjoint representation
\cite{Landsteiner:2001ky}. In cases where there are
more bosons than fermions in the adjoint (such as the present case),
a resummation of the series is impossible, since the series does not
converge. At present, there
is no known procedure to cure this pathology. In particular, the pure
non-commutative Yang-Mills theory seems to be sick.  

Let us now proceed to the non-planar corrections to the 3-point
vertices. The pattern is similar: whereas planar graphs take
the same form as in the commutative theory and they are divergent,
non-planar graphs are UV-finite, but singular at $\theta \rightarrow
0$. Non-vanishing diagrams involve $U(1)$ factor in at least one
of the external legs \cite{Armoni:2001xr}. The amplitude in the
 case of 3 external gluons,
when all gluons are in the $U(1)$ is
\beq
A^{\mu \nu \rho} _{(1-1-1)} =
{i64g^3 \sqrt {N/2} \over (4\pi)^2}  \left ( {\tilde p_1 ^\mu \tilde p_1
^\nu \tilde p_1 ^\rho \over \tilde p_1 ^4}
+{\tilde p_2 ^\mu \tilde p_2
^\nu \tilde p_2 ^\rho \over \tilde p_2 ^4}
+{\tilde p_3 ^\mu \tilde p_3
^\nu \tilde p_3 ^\rho \over \tilde p_3 ^4}
 \right ) \label{gtree1}.
\label{3point}
\eeq
We have ignored in this expression a factor $\cos {\tilde p}_1 p_2 /2$, 
which appears in previous calculations of the leading IR contribution 
to the 3-point function. The reason is that, in 
the approximation used to obtain \eqref{3point}, i.e. ${\tilde p}_i
p_j \ll 1$,
we cannot distinguish between $\cos {\tilde p}_1 p_2 /2$ and $1$.
We keep this convention in the following.
When one gluon is in the $U(1)$ and the two other gluons are in the $SU(N)$
the amplitude takes the form
\beq
A^{\mu \nu \rho} _{(1-N-N)} =
{i64g^3 \sqrt {N/2} \over (4\pi)^2} \ {\tilde p_1 ^\mu \tilde p_1
^\nu \tilde p_1 ^\rho \over \tilde p_1 ^4}
 , \label{gtree2}
\eeq
where $\tilde p_1$ is the momentum of the $U(1)$ field. Similarly,
 the amplitude for two external scalars and one gluon, all in
the $U(1)$, is
\beq
A^{\mu} _{(1-1-1)} =
{i32g^3 \sqrt {N/2} \over (4\pi)^2}  \left ( {\tilde p_1 ^\mu \over \tilde p_1 ^2}
+{\tilde p_2 ^\mu \over \tilde p_2 ^2}
+{\tilde p_3 ^\mu \over \tilde p_3 ^2}
 \right ) .
\eeq
In the case of two scalars and one gluon transforming in $U(1)$ and
$SU(N)$ the amplitude is the following
\beq
A^{\mu} _{(1-N-N)} =
{i32g^3 \sqrt {N/2} \over (4\pi)^2} \ {\tilde p_1 ^\mu \over \tilde p_1 ^2},
\eeq
where, again, $\tilde p_1$ is the momentum of the $U(1)$ field.

The information about the various non-planar diagrams can be
summarized in the following effective action
\bea 
& &
{\pi ^2 \over 2} S_I= g^2 \int d^4 p\ \left (2 \, {\tilde p^\mu \tilde p^\nu \over \tilde p^4}
\, {\rm tr}\, A_\mu (-p)\ {\rm tr}\, A_\nu (p)
+  {1\over \tilde p^2}\,
{\rm tr}\, \phi ^i (-p)\ {\rm tr}\, \phi ^i (p) \right ) \nonumber \\
& & 
+ \ \frac{i\, g^3}{(2 \pi)^4} \int d^4 p_1 \, d^4 p_2 \, d^4 p_3\, \delta (p_1 +p_2 +p_3) 
\times \nonumber \\
& &
\left \{\, 2 \; {\tilde p_1 ^\mu \tilde p_1
^\nu \tilde p_1 ^\rho \over \tilde p_1 ^4} \;
{\rm tr}\, A_\mu (p_1) \ {\rm tr}\, A_\nu (p_2) A_\rho (p_3) 
\nonumber \right . \\
& & 
\left . +\; {\tilde p_1 ^\mu \over
 \tilde p_1 ^2}
\left (  
{\rm tr}\, A_\mu (p_1) \ {\rm tr}\, \phi^i
 (p_2) \phi^i(p_3)  +
2 \; {\rm tr}\, \phi^i(p_1)\ {\rm tr}\, \phi^i(p_2) A^\mu (p_3)
\right ) \right \} . 
\label{action1} 
\eea

Apart from the terms which are summarized in the effective action
\eqref{action1}, there are other contributions which are less singular
when $\theta \rightarrow 0$. In contrast to the poles, these terms
(which as we shall see in a moment are log-like terms) do not cancel
even in the supersymmetric case, apart from the ${\cal N}=4$ SYM case
\cite{Matusis:2000jf}. These terms have a different Lorentz structure than
the poles and
they are all proportional to the one-loop beta function coefficient. Most
of our analysis of this part is based on \cite{Martin:2001bk} and \cite{Khoze:2001sy}.

The gluon propagator (for the $U(1)$ degrees of freedom) contains the
 following non-planar contribution
\beq
M^{\mu \nu} _{(1-1)} = -{26 g^2 N \over 3(4\pi)^2} 
(p^2 g^{\mu \nu} - p^\mu p^\nu) \log m^2 \tilde p ^2  ,
\eeq
where $m^2$ is an IR cut-off. We can think about it as
 a mass term for the scalars (and vectors), given via a Higgs
 mechanism. Similarly to the gluon, the correction to the scalar propagator is
\beq
M_{(1-1)}= -{26 g^2 N \over 3(4\pi)^2} p^2 \log m^2 \tilde p ^2 .
\eeq 

The subleading corrections to the 3-point vertices are the following:
for 3 gluons, all in the $U(1)$, we have
\beq
M^{\mu \nu \rho} _{(1-1-1)} = -{i26 g^3 \sqrt{N/2} \over 3(4 \pi)^2} \sin
({1\over 2} \tilde p_1 p_2) (\log m^2 \tilde p_1 ^2\ g ^{\nu \rho} p_1 ^\mu 
+{\rm perm.}),
\eeq
where 'perm.' means permutations of the three momenta and Lorentz indices
due to the symmetry of the amplitude. Similarly for the case
of 1 gluon in the $U(1)$ and 2 gluons are in the $SU(N)$ 
\beq
M^{\mu \nu \rho} _{(1-N-N)} = -{i26g^3 \sqrt{N/2} \over 3(4 \pi)^2} \sin
({1\over 2} \tilde p_1 p_2) (\log m^2 \tilde p_1 ^2\ g ^{\nu \rho} p_1 ^\mu 
+{\rm perm.}),
\eeq
where now $p_1$ is the momentum of the $U(1)$ gluon, and the
permutations are with respect to the 2 gluons in the $SU(N)$. 

In the case of amplitudes where there are two scalars and one gluon
we have 
 \beq
M^{\mu} _{(1-1-1)} = -{i26g^3 \sqrt{N/2} \over 3(4 \pi)^2} \sin
({1\over 2} \tilde p_1 p_2) (\log m^2 \tilde p_1 ^2\  p_1 ^\mu 
+{\rm perm.}),
\eeq
and the same expression for the $SU(N)-SU(N)-U(1)$ amplitude.

The log-like amplitudes can be summarized by the following effective action
\bea 
& &
-{24\pi ^2 \over 13} S_{II}= g^2 \int d^4 p\ \left ( (p^2 g^{\mu \nu} -
p^\mu p^\nu) \log m^2 \tilde p^2 ({\rm tr}\ A^\mu (-p)) ({\rm tr}\ A^\nu (p))
\right . \nonumber \\ 
& & 
\left . + p^2 \log m^2 \tilde p^2
({\rm tr}\ \phi ^i (-p) )({\rm tr}\ \phi ^i (p) \right ) \nonumber \\
& & 
+ {i g^3 \over (2\pi)^4} \int d^4 p_1\ d^4 p_2\ d^4 p_3\ \delta (p_1 +p_2 +p_3)\times \sin
({1\over 2} \tilde p_1 p_2) \times \log m^2 \tilde p_1 ^2 \ p_1^\mu \times \nonumber \\
& &
\left \{ ({\rm tr}\ A^\nu (p_1)) ({\rm tr}\ A^\mu (p_2) A^\nu (p_3) )
+
({\rm tr}\ \phi ^i (p_1)) ({\rm tr}\ A^\mu (p_2) \phi ^i (p_3) )
\right \} .
\label{action2} 
\eea

The actions \eqref{action1}\eqref{action2} are not gauge invariant. In order to have a
(non-commutative) gauge invariant action, higher order terms in
$A_\mu$ should be added. In the following sections
we will derive a manifestly gauge-invariant
action which includes \eqref{action1} and \eqref{action2} as part of it. 

\section{The Effective Action - \newline Field Theory Derivation}

\subsection{The Poles}

We will start by considering the pole-like IR-divergent 
contributions to the 2- and 3-point functions with only gluons 
as external legs. We observe that in both cases each vector field
$A_\mu(p_i)$ is contracted with $\tilde p^\mu=\theta^{\mu \nu}p_\nu$, 
where $p_\nu$ is the total momentum flowing on each trace operator.
This suggests that these terms are related to the simplest gauge-invariant
operators carrying non-zero momentum, the straight open Wilson 
line defined by \cite{Ishibashi:2000hs,Gross:2000ba}
\beq
W(p)\, = \, {\rm tr} \int d^4 x \; P_\ast \left(e^{i\, g \int_0^1 d \sigma 
\, {\tilde p}^\mu A_{\mu}(x+{\tilde p}\, \sigma)}\right) \ast e^{i p x} \, .
\label{w}
\eeq
Indeed, the gluon 2- and 3-point functions \eqref{action1} can be
obtained as the first 
terms in the expansion of the following gauge-invariant expression
\beq
S ^I _{eff}\, =\, \frac{2+N_s}{2 \pi^2} \int d^4 p \; W'(-p) f(\tilde p) \, 
W'(p) \, ,
\label{wilson}
\eeq
with $f(\tilde p)$ a function that tends in the IR to $1/\tilde p^4$; 
$N_s$ is the number of scalars in the adjoint representation
($N_s=6$ in the type 0 case). We denote by $W'(p)$ the Wilson
loop operator \eqref{w} once the $O(g^0)$ term has been subtracted
\beq
W'(p) =
 i\, g\, \tilde p^\mu {\rm tr} \, A_\mu (p) - 
g^2 \! \int \frac{d^4 l}{(2 \pi)^4} 
\frac{{\rm sin} \frac{{\tilde l} p}{2}}{{\tilde l} p}
\tilde p^\mu \tilde p^\nu 
{\rm tr}\, A_\mu (p\!-\!l) A_\nu(l) + ... \, . 
\label{exp}
\eeq
By inserting \eqref{exp} in \eqref{wilson}, we immediately recover
the gluon 2-point function. The expressions \eqref{gtree1},\eqref{gtree2} 
for the gluon 3-point function are valid in the 
limit $\tilde p_i p_j\!<\!<\!1$. In that limit ${\rm sin} 
\frac{{\tilde l} p}{2}/
\frac{{\tilde l} p}{2} \rightarrow 1$ and thus also the 3-point function 
is correctly obtained from \eqref{wilson}. This was to be expected since 
the IR divergent contribution to the 3-point function satisfies the Ward 
identity \cite{Martin:2001bk}.

For the pure $U(1)$ non-commutative theory, \eqref{wilson} can be obtained 
from a direct calculation of the 1-loop N-point functions. This will allow 
us to determine the function $f$ in \eqref{wilson}.
Due to the structure of the argument in the exponential of the Wilson loop, 
\eqref{wilson} contributes to the N-point function with terms proportional 
to $\tilde p^{\mu_1} ... \tilde p^{\mu_N}$. The N-point functions will have 
in general a complicated Lorentz index structure. However it is easy to 
isolate the terms of the mentioned form. They can only come from 
diagrams with 3-point vertices. Diagrams with 4-point vertices will give 
rise to a tensor structure containing $g^{\mu_i \mu_j}$, and therefore 
are not of the desired form.
We will like to point out that 
diagrams with 4-point vertices can produce as strong an IR divergence as 
those with only 3-point vertices. Indeed, the tadpole
induces a quadratic pole-like contribution to the 2-point function 
of the form $g^{\mu \nu}/{\tilde p}^2$. However the role of this term is to 
cancel a similar contribution coming from diagrams with 3-point vertices, 
and which would otherwise 
violate the Ward identity \cite{Matusis:2000jf}. The same applies to the 
3-point function. We will thus ignore diagrams with 4-point vertices when
analizing the leading UV/IR mixing effects.  

We will use the background field method in the 
following analysis; for the associated Feynman rules 
see \cite{Martin:2001bk}. The diagrams we are interested in are those 
depicted in fig.1. We have 
\bea
&&
(a)+(b)= \label{calc} \\
&& 
(-2 i g)^N \! \int \frac{d^4 l}{(2 \pi)^4} 
\prod_{i=1}^N \frac{(2 l + 2 p_1 +..+2p_{i-1}+
p_{i})_{\mu_i}}{(l+p_1+..+p_{i-1})^2} \,
{\rm sin} {\smallfrac{\tilde p_i (l+p_1+..+p_{i-1})}{2}}
\nonumber \, . 
\eea

\begin{figure}
  \begin{center}
\mbox{\kern-0.5cm
\epsfig{file=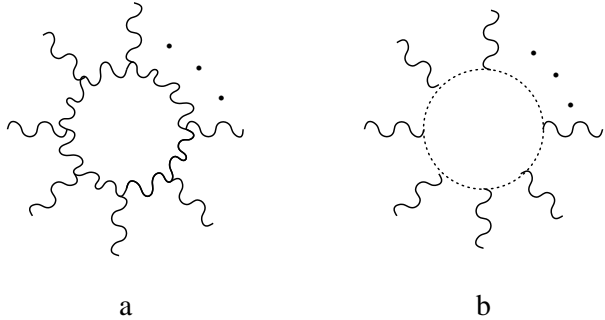,width=8.0true cm,angle=0}}
\label{diagram}
  \end{center}
\caption{Amplitudes containing terms 
$\sim \tilde p^{\mu_1} ... \tilde p^{\mu_N}$. Wavy lines refer to gauge 
bosons and doted lines to ghost. The end points of the external
 lines are background vector fields $B_\mu$.}
\end{figure}

\noindent As explained, we will disregard those 
parts of \eqref{calc} whose tensor
structure is such that they cannot contribute to \eqref{wilson}. 
This allows us to discard all the terms in the numerator proportional to
external momenta, and keep only $2l_{\mu_i}$ for each $i$. 
Expression \eqref{calc} then reduces to
\beq
(-2 i g)^N \,  \sum_{\nu_i} \, (-)^n
\int \frac{d^4 l}{(2 \pi)^4} \frac{l_{\mu_1}..l_{\mu_N} \, e^{-i \tilde p l 
- \frac{i}{2} \sum_{j<k} \tilde p_j 
p_k \nu_k}}{l^2 
(l+p_1)^2 ... (l+p_1+..+p_{N-1})^2} \, ,
\label{scalar2}
\eeq
where the summation on $\nu_i$, $\nu_i=\pm 1$ for $i=1,..,N$,
comes from expanding the sine. We have defined $p=\sum_i p_i \frac{1-\nu_i}{2}$ and $n= \sum_i
\frac{1-\nu_i}{2}$. We can interpret the N vertices as
twisted or untwisted depending if $\nu=-1$ or $1$ respectively. 
Thus $n$ is the number of twisted vertices and $p$ the total momentum
flowing in the twisted vertices. The $l_{\mu_i}$ in the numerator can now be
substituted by derivatives with respect to $\tilde p^{\mu}$
acting on the exponential. In order to simplify the analysis will we 
consider $\tilde p$ as an independent variable, and only in the end we 
will set ${\tilde p}^\mu\!=\!\theta^{\mu \nu} p_\nu$ with 
$p\!=\! \sum_i p_i \frac{1-\nu_i}{2}$. This allows us to bring the 
derivatives out of the integral, and rewrite \eqref{scalar2} as
\beq
(-2 i g)^N \, \partial_{\mu_1} ... \partial_{\mu_N} \sum_{\nu_i} \, (-)^n
\int \frac{d^4 l}{(2 \pi)^4} \frac{e^{-i \tilde p l - \frac{i}{2} 
\sum_{j<k} \tilde p_j 
p_k \nu_k}}{l^2 (l+p_1)^2 ... (l+p_1+..+p_{N-1})^2} \, .
\label{scalar22}
\eeq
The integral appearing in this expression coincides with that of
the N-point function of a non-commutative $\Phi^3$ theory and has been 
calculated in \cite{Kiem:2001pw,Kiem:2001dk} (see also
  \cite{Kiem:2001du} for a recent two-loops analysis). There, a small
 mass $m$ for the field $\Phi$ was introduced as an ordinary infrared
 regulator. The evaluation
of the previous integral gives \footnote{This result is
not affected by considering $p$ and $\tilde p$ as independent variables.}
\beq
J_{\ast n} (-p) \left(\frac{ \tilde p}{m}\right)^{\!\!N-2}\!\!\!\! \!\! 
K_{N-2}\,(m \tilde p) \, J_{\ast N-n}(p) \, ,
\label{scalar3}
\eeq
where $K_{N-2}$ are modified Bessel functions. We have denoted by 
$J_{\ast n}(-p)$ the kernel of the $\ast _n$-product 
defined in \cite{Liu:2000ad}, i.e. $J_{\ast n} (-p)\equiv 
J(-p_{r(1)},..,-p_{r(n)})$ where $p_{r(j)}$ are the $n$ momenta entering 
the twisted vertices and $p=\sum p_{r(j)}$. A comment is now in order.
Expression \eqref{scalar3} is not the complete answer, but
the leading term in the infrared. Subleading 
terms are suppressed by powers of ${\tilde p}^2 p_i p_j$ and therefore they 
do not give rise to infrared divergences for any $N$ 
\footnote{It is interesting that the subleading terms 
do not seem to have such a simple expression in terms of $\ast_n$ 
products as \eqref{scalar3}.}. In the following, we will keep in the 
evaluation of the N-point functions only the infrared-leading term. 
Then, \eqref{scalar22} reduces to 
\beq
\frac{1}{2 \pi^2}(-i g)^N \sum_{\nu_i} (-)^n
J_{\ast n} (-p) \left[ \partial_{\mu_1} ... \partial_{\mu_N} 
\left(\frac{ \tilde p}{m}\right)^{\!\!N-2}\!\!\!\! \!\! K_{N-2}\,(m \tilde p) \right] J_{\ast N-n}(p) \, .
\eeq
Using the properties of the modified Bessel functions, it is easy to 
see that the term in square brackets gives rise to a contribution of the 
form
\beq
(-)^N \, \tilde p^{\mu_1} ... \tilde p^{\mu_N} \; \frac{m^2}{\tilde p^2} \,
K_2(m \tilde p) \, .
\eeq
Adding up all such contributions to the effective action we get
\bea
S^I _{eff}& = &\frac{1}{2 \pi^2} \, \sum_{N=2}^\infty \; (i g)^N 
\int d^4 p \, \sum_{n=1}^{N-1} \, \frac{(-)^n}{n!(N-n)!} 
 \\ 
&&
\frac{m^2}{\tilde p^2} K_2(m \tilde p)\;
\tilde p^{\mu_1} ... \, \tilde p^{\mu_N}\, [A_{\mu_1}..A_{\mu_n}]_{\ast n} 
(-p) \; [A_{\mu_{n+1}}..A_{\mu_N}]_{\ast N-n} (p)
\, . \nonumber \\
\nonumber
\eea
This expression reproduces \eqref{wilson} by setting $2 f(\tilde p)=
\frac{m^2}{\tilde p^2} K_2(m \tilde p)$.
Although the $\ast _n$ also appear in the effective action of the
non-commutative $\Phi^3$ theory, they only combine to form
the scalar analog of Wilson loop operators in the limit of large
non-commutative parameter \cite{Kiem:2001dk}. On the contrary, the invariance
of the effective action with respect to gauge transformations of the 
background field suggests that, in gauge theories, Wilson loop operators 
will play an important role for 
any value of $\theta$. As a first example, a Wilson loop completion of 
the non-planar contributions to the $F^4$ terms in ${\cal N}=4$ gauge theory 
has been proposed in \cite{Liu:2000mj}. 
We have just seen that the puzzling pole-like divergent terms originating 
from UV/IR mixing are part of the simplest
gauge-invariant double-trace operator that can appear in the effective
action. We will leave for the next section the extension of the previous
considerations to gauge theories with adjoint matter.

\subsection{The Logarithms}

We would like to comment on the IR logarithmic-divergent
 terms arising from UV/IR mixing. As already mentioned, these
subleading contributions occur also in the supersymmetric case. We
suggest here a gauge invariant completion of the IR logarithmic
 divergent terms. This
 suggestion is not as rigorous as the derivation in the previous
 subsection, but our result is fixed by the requirement of gauge invariance.

It was shown in \cite{Martin:2001bk} that the logarithmic singularities
of the 2-, 3- and 4-point function of pure NC $U(1)$, in the limit
$|\tilde p_i| \sim |\tilde p_i + \tilde p_j| \sim \theta 
\Lambda_{IR}\rightarrow 0$,
combine into the following contribution to the effective action:
\beq
S^{II}_{eff} \, =\, \frac{1}{4} \, \beta_0 \, \rm{log} \, 
(\theta\Lambda_{IR})^2 
\int d^4 x \, F^{\mu \nu} F_{\mu \nu} \, ,
\label{lo}
\eeq 
with $\Lambda_{IR}$ an infrared cut-off and $\beta_0$ the coefficient
of the 1-loop beta function. It is tempting to propose the following 
gauge-invariant completion of \eqref{lo}, which generalizes to the
$U(N)$ case 
\beq
S^{II}_{eff}\, =\, \frac{1}{4} \, \beta_0  
\int d^4 p \, {\cal O}^{\mu \nu}(-p) K_0(m \tilde p)
{\cal O}_{\mu \nu} (p)\, ,
\label{loeff}
\eeq 
where the operator ${\cal O}_{\mu \nu}$ is defined by
\beq
{\cal O}_{\mu \nu} (p) = \rm{tr} \int d^4 x\ L_\ast \left( F_{\mu \nu} (x) \,
e^{i\, g \int_0^1 d \sigma 
\, {\tilde p}^\mu A_{\mu}(x+{\tilde p}\, \sigma)}\right) \ast e^{i p x}
\, .
\eeq
Following the notation of \cite{Liu:2000mj}, $L_\ast$ denotes integration 
of $F_{\mu \nu}$ along the open Wilson line together with path
ordering with respect of the $\ast$-product of all terms inside
the parenthesis. The action \eqref{loeff} reproduces the pure gluonic
log-like N-point functions \eqref{action2} in the small $m$ limit.

\section{The Effective Action via \newline Closed Strings Exchange}

The recent interest in the study of non-commutative field theories 
has been mainly motivated by their connection to string theory.
The world-volume coordinates of D-branes in the presence of a 
constant B-field background turn out to satisfy the relation 
$[x^\mu,x^\nu]=i\theta ^{\mu \nu}$, with $\theta^{\mu \nu}\!\sim\! 
1/B_{\mu \nu}$. 
As a consequence, the low energy theory on the brane is a 
non-commutative gauge theory.
In this section we would like to analyze \eqref{wilson} and
\eqref{loeff} (or \eqref{action1} and \eqref{action2}) 
from a string-inspired point of view. 
In a series of recent papers it has been shown that closed string 
modes couple to non-commutative D-branes through open Wilson line
operators \cite{Liu:2000mj,Das:2001ur,Okawa:2001sh,Liu:2001ps}. 
This result was obtained by evaluating the 
disk amplitude between a closed string and open string modes.

Let us consider the annulus diagram with
boundaries on non-commutative D-branes as in fig.2. 
\begin{figure}
  \begin{center}
\mbox{\kern-0.5cm
\epsfig{file=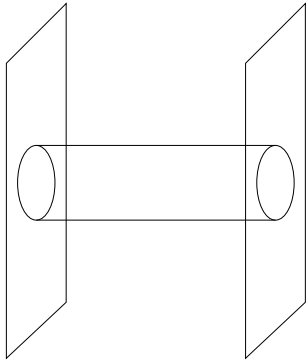,width=4.0true cm,angle=0}}
\label{annulus}
  \end{center}
\caption{The annulus amplitude.}
\end{figure}
It can be seen as a loop of open strings or a tree level exchange of 
closed strings. In the limit of a large cylinder the closed string channel 
picture is more adequate since the annulus
diagram factorizes to closed string insertions on a disk connected
by a closed string propagator \cite{Douglas:1997yp}. In the opposite limit 
of a small cylinder, the exchange of the lowest open string modes dominates.
This provides the field theory limit, and the annulus amplitude reproduces
the 1-loop field theory effective action. 
Thus in general we could expect in the field theory effective action
more complicated contributions than \eqref{wilson} and \eqref{loeff},
which structurally are reminiscent of a closed string exchange.
Notice that a similar structure was proposed
as the gauge-invariant completion of the non-planar $F^4$ terms in
the effective action of ${\cal N}=4$ non-commutative Yang-Mills
\cite{Liu:2000ad,Liu:2000mj}. In that case, the function $f$ had the 
interpretation of a closed
string propagator in type II string theory. This however comes as no surprise
since the $F^4$ terms in the maximally supersymmetric case are protected
by non-renormalization theorems \cite{Douglas:1997yp}. 
In contrast, it is
remarkable that \eqref{wilson} emerges in a non-supersymmetric theory.
We will show below that the function $f$ appearing in the IR-divergent terms
can also be directly related to a closed string propagator.   

In the rest of this section we will consider type 0 string theory.
This theory can be obtained as a world-sheet orbifold of type
II, which projects out space-time fermions. It contains a closed
string tachyon arising from the twisted sectors. There are however
no open string tachyons on D-branes in type 0 theory. This makes it 
especially adequate for our considerations. We will work with
the gauge theory on $N$ electric D3-branes. It is given by the dimensional 
reduction of 
pure Yang-Mills in 10 dimensions, i.e. gauge fields plus 6 scalars in
the adjoint representation.

We start by analysing which closed string modes couple to the open Wilson line 
operator \eqref{w}. The first candidate is the type 0 tachyon. In the 
absence of B-field and at leading order in $\alpha'$, it couples to 
the brane tension as \cite{Klebanov:1999yy,Garousi:1999fu}
\beq
\frac{N}{4\, (2 \pi \alpha')^2} \, .
\label{ten}
\eeq
Following the same analysis done for 
bosonic and type II string theory \cite{Liu:2000mj,Okawa:2001sh}, it is easy to see
that the trivial field theory operator \eqref{ten} gets promoted to
an open Wilson loop in the presence of a B-field. The coupling
of the type 0 tachyon to the D-brane field theory at leading
$\alpha'$ order is described by
\beq
S^I= \frac{\kappa_{10}}{g_{YM}^2} \int \frac{d^{10} P}{(2 \pi)^{10}}
\, \sqrt{ \rm{det}\, G} \  T(P) \ {\cal O}(-P)  \, ,
\eeq
where $G$ is the open string metric and 
\beq
{\cal O}(P) = \frac{1}{4\,(2 \pi \alpha')^2} \, {\rm tr} \int d^4 x \, W(x,C) 
\ast e^{i p x}  \, .
\label{tac}
\eeq
We denote by $P_M$ the 10-dimensional momentum, $p_\mu$ the momentum along the 4-dimensional
world-volume of the D3-brane and ${p_\perp}_i$ the momentum in the transverse directions.
In the previous expressions $W(x,C)$ is a generalization of \eqref{w} which involves
the transverse scalars
\beq
W(x,C)=  P_\ast \left(e^{i\, g \int_0^1 d \sigma 
\, {\tilde p}^\mu A_{\mu}(x+{\tilde p}\sigma) + y_i \phi^i(x+ {\tilde p} \sigma)} \right) \, ,
\eeq
where we have defined $y_i=2 \pi \alpha' {p_\perp}_i$ and $\phi^i= X^i/2 \pi \alpha'$ for $i=1,...,6$,
which provides the correct normalization for the field theory scalar fields.

The on-shell condition for the type 0 tachyon is $P_M g^{MN} P_N = -2/\alpha'$,
with $g$ the closed string metric. Closed and open string 
metrics are related by $g^{-1}=G^{-1}\!-\!\theta G \theta/(2 \pi \alpha')^2$
\cite{Seiberg:1999vs}. In the Seiberg-Witten limit, i.e. 
$\alpha'\!\rightarrow\!0$ keeping $G$ and $\theta$ fixed, the on-shell 
condition becomes
\cite{Liu:2000mj,Okawa:2001sh}
\beq
{\tilde p}^2 + y^2 = 8 \pi^2 \alpha' \, .
\label{ms}
\eeq
The closed string mass is a subleading effect with respect to the momentum 
in the non-commutative directions in the Seiberg-Witten limit. 
In spite of that, it will be crucial in the following to keep its 
contribution to the mass-shell condition.
We want to analyse how the tachyon exchange contributes to the
annulus amplitude. For two D3-branes separated by a distance 
$r$ we obtain    
\beq
S^I_{e\!f\!f}=\frac{\kappa_{10}^2}{g_{YM}^4} \int \frac{d^{10} P}{(2 \pi)^{10}}
\, \frac{{\rm det}\, G}{\sqrt{{\rm det} \, g}}\ 
{\cal O}(P) 
\, {\cal O}(-P) \ \frac{e^{i p_\perp r}}{\frac{M^2}{(2 \pi \alpha')^2}+p_\perp^2} \, .
\label{seff}
\eeq
The quantity $M^2/(2 \pi \alpha')^2$ is the effective mass of the closed 
string tachyon propagating in the six transverse dimensions; 
from \eqref{ms} $M^2={\tilde p}^2 - 8 \pi^2 \alpha'$.

In order to make contact with the previous section we will first consider 
the dependence of ${\cal O}$ on the gauge fields only. Then 
${\cal O} \!= \frac{1}{(4 \pi \alpha')^2} \!W(p)$,
with $W(p)$ given by \eqref{w}. Using (see for example \cite{Liu:2000mj}) 
\beq
{\kappa_{10}^2 \over g_{YM}^4}=\pi (2 \pi \alpha')^4 {\sqrt{ {\rm det}\, g}
\over \sqrt{ {\rm det} \, G}} \, ,
\eeq 
and defining $m=r/2 \pi \alpha'$, the previous expression can be rewritten as
\beq
S^I_{e\!f\!f}=\frac{\pi}{(4 \pi \alpha')^4} \int \frac{d^4 p}{(2
\pi)^4} \, \sqrt{{\rm det} \, G} \ {\rm tr}\, W(p)\ {\rm tr}\, W(-p) \ 
G(p) \, , 
\label{effdiv}
\eeq 
where
\beq
G(p)=\int \frac{d^6 y}{(2 \pi)^6} \frac{e^{i y m}}{M^2 + y^2}=\frac{1}
{(2 \pi)^3} \frac{M ^2}{m^2} K_2(m M) \, .
\label{prop}
\eeq 
$G(p)$ represents the closed string propagator in the transverse
dimensions, rescaled appropriately to the field theory limit. Indeed, 
it is finite in the limit $\alpha' \! \rightarrow \! 0$. 
However \eqref{effdiv} diverges in this limit due to the $O(\alpha'^{-2})$ 
dependence of the brane tension to which the tachyon couples. We 
can define a finite contribution to \eqref{effdiv} by expanding $G(p)$
to $O(\alpha'^4)$, using the explicit dependence of $M^2$ on $\alpha'$. 
We then obtain a contribution to the field 
theory effective action of the form \eqref{wilson}, with
\beq
f(p)\equiv  G(p)|_{\alpha'^4}= c \, \frac{m^2}{\tilde p^2} K_2(m \tilde p) 
\label{f}
\, , 
\eeq
with $c\!=\!\frac{4 \pi^5}{3}$. This agrees with the result derived from 
the field theory calculation, up to a global coeficient. We will comment 
on this below. 
 
Notice that in order to obtain the IR divergent terms from the string
exchange, it was essential that the field theory operator that couples
to the tachyon carries negative powers of $\alpha'$. The reason for
this is that at ${\cal O}(\alpha'^0)$, $G\sim 1/m^4$ as 
${\tilde p} \rightarrow 0$.
Such a term is related to the ordinary infrared problems of a
field theory with massless degrees of freedom. However, remarkably,
$G(p)$ contains information about the new divergences 
due to UV/IR mixing effects in non-commutative field theories
when expanded to higher orders in $\alpha'$. We have analyzed above the
coupling of the type 0 tachyon to the D-brane field theory at 
leading order in $\alpha'$. At ${\cal O}(\alpha'^0)$ it couples to 
the field theory operators ${\rm tr} F^2$ and $(D \phi^i)^2$ 
\cite{Klebanov:1999yy,Garousi:1999fu}. For the reasons just exposed, the
coupling of the tachyon to these operators would contribute non-singular 
terms in the effective action and therefore we will not consider them.

Equation \eqref{effdiv} differs from \eqref{wilson}, as we
subtracted the '1' from the open Wilson line in \eqref{wilson}. The
'1' in coordinate space is in fact $\delta(p)$, as we work in Fourier space,
 and therefore this difference
 affects only the $p_\mu=0$
component of $W$. We would like to stress that the expansion of 
$G(p)$ in $\alpha'$ powers requires that ${\tilde p}$ is non zero. At ${\tilde p}=0$ and
in the limit $\alpha' \rightarrow 0$, the string propagator
is $G \sim 1/m^4$. The associated contribution to the effective 
action is proportional to
\beq
{1 \over (\alpha' m)^4}\; \delta^{(4)}(0) \; \rightarrow \;
\Lambda^4 \int d^4 x \, ,
\eeq   
where $1/(\alpha' m)\sim \Lambda$ can be interpreted as a field theory 
scale. Therefore the difference between \eqref{effdiv} and \eqref{wilson}
reflects the vacuum energy of the gauge theory, which is taken 
into account in the string theory calculation. Once this infinity
is substracted, the string exchange just reproduces the field theory
result \eqref{wilson}.

We will now show that \eqref{seff} can also reproduce the pole-like 
divergent terms associated with the adjoint scalars. Expanding ${\cal O}$ 
to linear order in the fields, we obtain the following contribution 
involving the adjoint scalars
\beq
S'^I_{e\!f\!f} \sim \int \frac{d^4 p}{(2 \pi)^4}\,\sqrt{{\rm det} \, G}\, {\rm tr}\,
\phi_i(p)\ {\rm tr}\, \phi_j(-p) \ f_{ij}(p) \, , 
\label{effdivs}
\eeq 
where
\bea
f_{ij}&=& \left. \int \frac{d^6 y}{(2 \pi)^6} \frac{e^{i y m}\, y_i 
y_j}{M^2 + y^2}\, \right|_{\alpha'^4}= \nonumber \\
&&
=- \partial_{m_i} \partial_{m_j} G(p) = c \left( 
\delta_{ij} \frac{m}{\tilde p} K_1(m \tilde p) + 
m_i m_j K_0(m \tilde p) \right) \, \label{fij}.
\eea
The first term in \eqref{fij} leads to the action  
\beq
S'^I_{e\!f\!f} \sim \int \frac{d^4 p}{(2 \pi)^4} \, \sqrt{{\rm det} \, G}\, {\rm tr}\,
\phi_i(p)\ {\rm tr}\, \phi_i(-p) \, {m\over \tilde p} K_1(m\tilde p) \, , 
\label{effdivs1}
\eeq
which corresponds, in the $m\tilde p \rightarrow 0$ limit, to the pole-like
contribution in the effective action of the scalars \eqref{action1}.
The second term in \eqref{fij} yields a $m^2 \log$ contribution which
vanishes when $m\rightarrow 0$. Notice that while 
$f$ in \eqref{f} tends to $2c/{\tilde p}^4$ in the infrared limit, 
$f_{ij}$ tends to $c/{\tilde p}^2$. This reproduces the relative factor 
of two between the pole-like contributions to the propagator
of the gauge field and adjoint scalars, eq.\eqref{vector},\eqref{scalar}.
The same applies to the linear poles of the 3-point functions. Therefore 
the gauge invariant expression \eqref{effdiv}, defined such that 
we only keep the finite terms in the $\alpha' \rightarrow 0$ limit, 
accounts for all pole-like divergent terms of the field theory
up to a global coefficient.

The discrepancy in the global coefficient can be related to the fact
that not only the tachyon, but also massive scalar closed strings 
couple to the brane tension. In the Seiberg-Witten limit 
these contributions are of the same form as that of the tachyon,
since momentum in the non-commutative directions dominates
over the oscillator mass. Thus they will renormalize 
the overall coefficient in front of the effective action. 
To summarize, we have seen that the gauge invariant effective action 
containing the infrared poles can be directly related to a closed 
string exchange between D-branes. It is tempting to think of this as 
the exchange of an ``effective closed string mode''. Remarkably,
among the original closed string modes that 
contribute to this effect is the tachyon mode.

We will briefly address the log like contributions
which appear also in the supersymmetric field theory \eqref{action2}. 
Consider a two-form (denoted by $M_{M N}$) closed string which couples 
to the operator ${\cal O}^{M N}$ (separated into 4d and 6d indices):     
\beq
S^{II}= \frac{\kappa_{10}}{g_{YM}^2} \int \frac{d^{10} P}{(2 \pi)^{10}}
\sqrt{ \rm{det}\, G} \;
 \left ( M_{\mu \nu} (P) {\cal O} ^{\mu \nu} (-P)+
  M_{\mu i} (P) {\cal O} ^{\mu i} (-P) \right ) \, ,
\eeq
with
\bea
& &
{\cal O}^{\mu \nu}(P) = \frac{1}{2 \pi \alpha'} {\rm tr} \int d^4 x
\, L_\ast (F ^{\mu \nu} W(x,C)) 
\ast e^{i p x}  \, , \nonumber \\
& &
{\cal O}^{\mu i}(P) = \frac{1}{2 \pi \alpha'} {\rm tr} \int d^4 x
\, L_\ast (D ^\mu \phi ^i W(x,C)) 
\ast e^{i p x}  \, .
 \label{2formac} 
\eea
Repeating the same steps as for the tachyon field we can write the
effective action due to an exchange of a massive 2-form as
\beq
S^{II}_{e\!f\!f} \sim
\int \frac{d^4p d^6y}{(2 \pi)^{10}}\, \sqrt{{\rm det}\, 
G} \ {\cal O}^{MN}(p,y) 
\, {\cal O}_{MN}(-p,-y) \frac{e^{i y m}}{M^2+y^2} \, ,
\label{seff2}
\eeq
with $M^2 = \tilde p ^2 + 8\pi^2 l \alpha' $ and $l$ some positive integer
number which corresponds to the string excitation number. For simplicity
let us set the adjoint scalar fields to zero in $W(x,C)$, which does not
affect gauge invariance. We get then 
\beq
S^{II}_{e\!f\!f} \sim \int \frac{d^4 p}{(2 \pi)^4}
 \left ( {\cal O} _{\mu \nu}(p) {\cal O} ^{\mu \nu} (-p)+
 {\cal O} _{\mu i}(p) {\cal O} ^{\mu i} (-p) \right ) G(p) \, , 
\label{effdiv2}
\eeq 
but now we should simply keep the terms in $G(p)$ which are $O(\alpha'^{2})$. This yields
\beq
  G(p)|_{\alpha'^2}\sim K_0(m \tilde p) \, .
\eeq
which reproduces the action \eqref{loeff} and in addition the log 
like pieces of the scalars \eqref{action2}.

Thus, we have shown that the logarithmic like ($K_0$, in fact)
contribution to effective action can be understood from massive 
2-form closed strings exchange. Note that the massless NS-NS 2 form does 
not contribute here. Only massive modes. Another comment is that we
could not reproduce the overall factor in front of the effective
action, $\beta _0$. The understanding of the overall factor, from the
string theory point of view, is equivalent to the understanding of the
weight of each individual massive string in the coupling to the
operator $F^{MN}$ on the brane. We will not address this problem here.

\section{Discussion} 
 In this work we have discussed UV/IR effects in a non-supersymmetric 
gauge theory. Our main results are the effective actions
\eqref{mainresult1} and \eqref{mainresult2}. 
These actions incorporate the two kinds of non-planar singular amplitudes:
the poles and the logs.

The log like contributions exist also in the supersymmetric theory,
apart from the ${\cal N}=4$ SYM theory. The picture that emerges from
our work is that one can understand these effects as due to an
exchange of massive two-form closed strings which couple 
to the operator ${\rm tr}\ F_{\mu \nu}$.

The more interesting contributions are the poles. These poles 
cancel in the supersymmetric gauge theory. Our picture here is that
these terms can be understood as due to an exchange of a tachyon and
massive scalar closed strings that couple to the brane tension. In the
superstring theory there is no tachyon. Moreover, the
contributions from the NS-NS sector cancel the contributions from the 
R-R sector and this is our explanation of why we do not see such effects
in the (super-)gauge theory side.

The (partial) contribution of the closed string tachyon to the
tachyonic instabilities of the non-commutative theory suggests
that the two phenomena are related. Indeed, it is true that the poles are
also due to massive closed strings,
since in the Seiberg-Witten limit all the massive tower
contributes similarly to the exchange between the
D-branes \cite{Arcioni:2000bz}. Therefore we do not argue that the 
tachyon in the field
theory has a one to one correspondence with the closed string tachyon. The
relation is more indirect. We have not found any example of a
non-supersymmetric string theory with a NS-NS two form which does
not contain a closed string tachyon (or a tree level open string tachyon).
It is possible to construct a non-tachyonic non-supersymmetric 
string theory \cite{Sagnotti:1995ga,Sagnotti:1997qj} by using a special orientifold, but the
orientifold removes the NS-NS two form from the spectrum. In addition,
in non-supersymmetric string theories, such as strings on orbifold
singularities, there is always a tachyon
in the twisted sector. In these cases the non-commutative field
theory on the brane is 'tachyonic'. Namely, in all these
 constructions there are
more bosons in the adjoint representation than fermions \footnote{A.A.
would like to thank Rodolfo Russo for discussions on this issue.}.
Finally, the effective action \eqref{mainresult2} is not tachyonic and
indeed it can be understood as due to massive 2-forms exchange (no
closed string tachyon is involved in this case).   
These observations support our claim about a relation between the closed
string tachyon and the generated tachyon on the brane. We suggest that
the closed string tachyon that couples non-trivially to the brane (in
contrast to the commutative theory), is behind the instabilities 
in the field theory. This point of view is somewhat similar in spirit
to \cite{Klebanov:1999um}, where it is was argued that a field theory
which is holographically dual to a tachyonic string theory should
suffer from instabilities. 

In the light of our picture, we would like to address the problem of the
stability of the non-commutative non-supersymmetric Yang-Mills theory.
Since this theory is tachyonic, similarly to type 0 string theory, we
suggest that the consistency issue is related to the fate of the closed
string tachyon. Type 0 string theory
might be consistent if tachyon condensation occurs (for concrete
suggestions see \cite{Bergman:1999km,Costa:2001nw,Adams:2001sv}).
In particular, the true vacuum of the type 0 string might be
supersymmetric ! It is tempting to suggest that if this
is the case, there will be examples of non-commutative non-supersymmetric 
gauge theories which are consistent and that \eqref{disp} is a consequence 
of the expansion around the perturbative vacuum.

\Acknowledgements
We would like to thank Luis Alvarez-Gaume, Carlo Angelantonj, Jose
Barbon, Chong-Sun Chu, Yaron Oz and
Rodolfo Russo for discussions. Part of the work of A.A. was done at
the Theoretical Physics Centre of the Ecole Polytechnique. A.A. wishes
 to take this opportunity to thank all members of the String Theory
 Group there.


\begin{thebibliography}{99}

\bibitem{Connes:1998cr}
A.~Connes, M.~R.~Douglas and A.~Schwarz,
``Noncommutative geometry and matrix theory: Compactification on tori,''
JHEP {\bf 9802}, 003 (1998)
[hep-th/9711162].

\bibitem{Seiberg:1999vs}
N.~Seiberg and E.~Witten,
``String theory and noncommutative geometry,''
JHEP {\bf 9909}, 032 (1999)
[hep-th/9908142].

\bibitem{Gonzalez-Arroyo:1982ub}
A.~Gonzalez-Arroyo and M.~Okawa,
``A Twisted Model For Large N Lattice Gauge Theory,''
Phys.\ Lett.\ B {\bf 120}, 174 (1983).

\bibitem{Gonzalez-Arroyo:1982hz}
A.~Gonzalez-Arroyo and M.~Okawa,
``The Twisted Eguchi-Kawai Model: A Reduced Model For Large N Lattice Gauge Theory,''
Phys.\ Rev.\ D {\bf 27}, 2397 (1983).

\bibitem{Filk:1996dm}
T.~Filk,
``Divergencies in a field theory on quantum space,''
Phys.\ Lett.\ B {\bf 376} (1996) 53.

\bibitem{Minwalla:2000px}
S.~Minwalla, M.~Van Raamsdonk and N.~Seiberg,
``Noncommutative perturbative dynamics,''
JHEP {\bf 0002}, 020 (2000)
[hep-th/9912072].

\bibitem{Matusis:2000jf}
A.~Matusis, L.~Susskind and N.~Toumbas,
``The IR/UV connection in the non-commutative gauge theories,''
JHEP {\bf 0012}, 002 (2000)
[hep-th/0002075].

\bibitem{Arcioni:2000bz}
G.~Arcioni, J.~L.~Barbon, J.~Gomis and M.~A.~Vazquez-Mozo,
``On the stringy nature of winding modes in noncommutative thermal field  theories,''
JHEP {\bf 0006}, 038 (2000)
[hep-th/0004080].

\bibitem{Arefeva:2000uu}
I.~Y.~Arefeva, D.~M.~Belov, A.~S.~Koshelev and O.~A.~Rychkov,
``Renormalizability and UV/IR mixing in noncommutative theories with scalar fields,''
Phys.\ Lett.\ B {\bf 487}, 357 (2000).

\bibitem{Micu:2001hp}
A.~Micu and M.~M.~Sheikh Jabbari,
``Noncommutative $\phi^4$ theory at two loops,''
JHEP {\bf 0101}, 025 (2001).


\bibitem{Chepelev:2001hm}
I.~Chepelev and R.~Roiban,
``Convergence theorem for non-commutative Feynman graphs and  renormalization,''
JHEP {\bf 0103}, 001 (2001)
[hep-th/0008090].

\bibitem{Griguolo:2001ez}
L.~Griguolo and M.~Pietroni,
``Wilsonian renormalization group and the non-commutative IR/UV  connection,''
JHEP {\bf 0105}, 032 (2001)
[hep-th/0104217].

\bibitem{Kinar:2001yk}
Y.~Kinar, G.~Lifschytz and J.~Sonnenschein,
``UV/IR connection: A matrix perspective,''
JHEP {\bf 0108}, 001 (2001)
[hep-th/0105089].

\bibitem{Hayakawa:1999zf}
M.~Hayakawa,
``Perturbative analysis on infrared and ultraviolet aspects of  noncommutative QED on $R^4$,''
hep-th/9912167.

\bibitem{Bilal:2000bk}
A.~Bilal, C.~Chu and R.~Russo,
``String theory and noncommutative field theories at one loop,''
Nucl.\ Phys.\ B {\bf 582}, 65 (2000)
[hep-th/0003180].


\bibitem{Armoni:2001xr}
A.~Armoni,
``Comments on perturbative dynamics of non-commutative Yang-Mills theory,''
Nucl.\ Phys.\ B {\bf 593}, 229 (2001)
[hep-th/0005208].

\bibitem{Landsteiner:2000bw}
K.~Landsteiner, E.~Lopez and M.~H.~Tytgat,
``Excitations in hot non-commutative theories,''
JHEP {\bf 0009}, 027 (2000)
[hep-th/0006210].

\bibitem{Martin:2001bk}
C.~P.~Martin and F.~Ruiz Ruiz,
``Paramagnetic dominance, the sign of the beta function and UV/IR mixing  in non-commutative U(1),''
Nucl.\ Phys.\ B {\bf 597}, 197 (2001)
[hep-th/0007131].

\bibitem{Pernici:2001va}
M.~Pernici, A.~Santambrogio and D.~Zanon,
``The one-loop effective action of noncommutative N = 4 super Yang-Mills  is gauge invariant,''
Phys.\ Lett.\ B {\bf 504}, 131 (2001)
[hep-th/0011140].

\bibitem{Khoze:2001sy}
V.~V.~Khoze and G.~Travaglini,
``Wilsonian effective actions and the IR/UV mixing in noncommutative  gauge theories,''
JHEP {\bf 0101}, 026 (2001)
[hep-th/0011218].

\bibitem{Zanon:2001nq}
D.~Zanon,
``Noncommutative N = 1,2 super U(N) Yang-Mills: UV/IR mixing and  effective action results at one loop,''
Phys.\ Lett.\ B {\bf 502}, 265 (2001)
[hep-th/0012009].

\bibitem{Ruiz:2001hu}
F.~R.~Ruiz,
``Gauge-fixing independence of IR divergences in non-commutative U(1),  perturbative tachyonic instabilities and supersymmetry,''
Phys.\ Lett.\ B {\bf 502}, 274 (2001)
[hep-th/0012171].

\bibitem{Landsteiner:2001ky}
K.~Landsteiner, E.~Lopez and M.~H.~Tytgat,
``Instability of non-commutative SYM theories at finite temperature,''
JHEP {\bf 0106}, 055 (2001)
[hep-th/0104133].


\bibitem{Armoni:2001md}
A.~Armoni and R.~Russo,
``Non-commutative gauge theories and the cosmological constant,''
hep-th/0106189.

\bibitem{Douglas:2001ba}
M.~R.~Douglas and N.~A.~Nekrasov,
``Noncommutative field theory,''
hep-th/0106048.

\bibitem{Szabo:2001kg}
R.~J.~Szabo,
``Quantum Field Theory on Noncommutative Spaces,''
hep-th/0109162.

\bibitem{Ishibashi:2000hs}
N.~Ishibashi, S.~Iso, H.~Kawai and Y.~Kitazawa,
Nucl.\ Phys.\ B {\bf 573}, 573 (2000)
[hep-th/9910004].

\bibitem{Gross:2000ba}
D.~J.~Gross, A.~Hashimoto and N.~Itzhaki,
``Observables of non-commutative gauge theories,''
hep-th/0008075.

\bibitem{MVR}
M.~Van Raamsdonk,
``The Meaning of Infrared Singularities in Noncommutative Gauge
Theories,'' hep-th/0110093.

\bibitem{Kiem:2001pw}
Y.~Kiem, S.~J.~Rey, H.~T.~Sato and J.~T.~Yee,
``Open Wilson lines and generalized star product in nocommutative
scalar  field theories,'' hep-th/0106121.

\bibitem{Kiem:2001dk}
Y.~Kiem, S.~Rey, H.~Sato and J.~Yee,
``Anatomy of one-loop effective action in noncommutative scalar field
theories,'' hep-th/0107106.

\bibitem{Kiem:2001du}
Y.~Kiem, S.~S.~Kim, S.~J.~Rey and H.~T.~Sato,
``Anatomy of Two-Loop Effective Action in Noncommutative Field 
Theories,'' hep-th/0110066.

\bibitem{Liu:2000ad}
H.~Liu and J.~Michelson,
``*-Trek: The one-loop N = 4 noncommutative SYM action,''
hep-th/0008205.

\bibitem{Liu:2000mj}
H.~Liu,
``*-Trek II: *n operations, open Wilson lines and the Seiberg-Witten  map,''
hep-th/0011125.

\bibitem{Das:2001ur}
S.~R.~Das and S.~P.~Trivedi,
``Supergravity couplings to noncommutative branes, open Wilson lines and  generalized star products,''
JHEP {\bf 0102}, 046 (2001)
[hep-th/0011131].

\bibitem{Okawa:2001sh}
Y.~Okawa and H.~Ooguri,
``How noncommutative gauge theories couple to gravity,''
Nucl.\ Phys.\ B {\bf 599}, 55 (2001)
[hep-th/0012218].

\bibitem{Liu:2001ps}
H.~Liu and J.~Michelson,
``Supergravity couplings of noncommutative D-branes,''
hep-th/0101016.

\bibitem{Douglas:1997yp}
M.~R.~Douglas, D.~Kabat, P.~Pouliot and S.~H.~Shenker,
``D-branes and short distances in string theory,''
Nucl.\ Phys.\ B {\bf 485}, 85 (1997)
[hep-th/9608024].

\bibitem{Klebanov:1999yy}
I.~R.~Klebanov and A.~A.~Tseytlin,
``Asymptotic freedom and infrared behavior in the type 0 string approach  to gauge theory,''
Nucl.\ Phys.\ B {\bf 547}, 143 (1999)
[hep-th/9812089].

\bibitem{Garousi:1999fu}
M.~R.~Garousi,
``String scattering from D-branes in type 0 theories,''
Nucl.\ Phys.\ B {\bf 550}, 225 (1999)
[hep-th/9901085].

\bibitem{Sagnotti:1995ga}
A.~Sagnotti,
``Some properties of open string theories,''
hep-th/9509080.

\bibitem{Sagnotti:1997qj}
A.~Sagnotti,
``Surprises in open-string perturbation theory,''
Nucl.\ Phys.\ Proc.\ Suppl.\  {\bf 56B}, 332 (1997)
[hep-th/9702093].

\bibitem{Klebanov:1999um}
I.~R.~Klebanov,
``Tachyon stabilization in the AdS/CFT correspondence,''
Phys.\ Lett.\ B {\bf 466}, 166 (1999)
[hep-th/9906220].

\bibitem{Bergman:1999km}
O.~Bergman and M.~R.~Gaberdiel,
``Dualities of type 0 strings,''
JHEP {\bf 9907}, 022 (1999)
[hep-th/9906055].

\bibitem{Costa:2001nw}
M.~S.~Costa and M.~Gutperle,
``The Kaluza-Klein Melvin solution in M-theory,''
JHEP {\bf 0103}, 027 (2001)
[hep-th/0012072].

\bibitem{Adams:2001sv}
A.~Adams, J.~Polchinski and E.~Silverstein,
``Don't panic! Closed string tachyons in ALE space-times,''
hep-th/0108075.
\end{thebibliography}
\end{document}